\documentclass[iop,numberedappendix]{emulateapj} \def \apjloutput {}

\usepackage{xcolor}
\usepackage{hyperref}
\hypersetup{colorlinks,citecolor=blue}
\usepackage{graphicx}
\usepackage{amssymb}
\usepackage{amsmath}
\usepackage{rotating}
\usepackage{float}
\usepackage{changepage}

\def\HST{\textit{HST}}

\def\wise{\textit{WISE}}
\def\WISE{\textit{WISE}}

\def\herschel{\textit{Herschel}}
\def\Herschel{\textit{Herschel}}
\def\W1{\textit{W1}}
\def\W2{\textit{W2}}
\def\W3{\textit{W3}}
\def\W4{\textit{W4}}
\def\Htwo{H$_{\rm 2}$}
\def\HeII{He\,{\sc ii}}
\def\MgII{Mg\,{\sc ii}}
\def\CIV{C\,{\sc iv}}
\def\CII{[C\,{\sc ii}]}
\def\FeII{Fe\,{\sc ii}}
\def\Ha{H{$\alpha$}}
\def\Hb{H{$\beta$}}
\def\kms{$\rm km\,s^{-1}$}
\def\w2246m05{W2246$-$0526}

\def\eEdd{$\lambda_{\rm Edd}$}
\def\hotdog{Hot DOG}
\def\hotdogs{Hot DOGs}

\ifx \apjloutput \undefined
\else
\slugcomment{{Accepted by the Astrophysical Journal}}
\fi

\begin{document}

\title{
Super-Eddington Accretion in the \WISE-Selected\\ 
Extremely Luminous Infrared Galaxy W2246$-$0526
}

\author{
Chao-Wei Tsai\altaffilmark{1},
Peter R. M. Eisenhardt\altaffilmark{2}, 
Hyunsung D. Jun\altaffilmark{3},
Jingwen Wu\altaffilmark{4}, \\
Roberto J. Assef\altaffilmark{5},
Andrew W. Blain\altaffilmark{6}, 
Tanio Diaz-Santos\altaffilmark{5},
Suzy F. Jones\altaffilmark{7}, \\
Daniel Stern\altaffilmark{2},
Edward L. Wright\altaffilmark{1}, and
Sherry C. C. Yeh\altaffilmark{8}
}

\altaffiltext{1}{Department of Physics and Astronomy, UCLA, Los Angeles, CA 90095-1547, USA; [email: cwtsai@astro.ucla.edu]}
\altaffiltext{2}{Jet Propulsion Laboratory, California Institute of Technology, Pasadena, CA 91109, USA}
\altaffiltext{3}{School of Physics, Korea Institute for Advanced Study, Seoul 02455, Korea}
\altaffiltext{4}{National Astronomical Observatories, Chinese Academy of Sciences, Beijing, 100012, China}
\altaffiltext{5}{N\'ucleo de Astronom\'ia de la Facultad de Ingenier\'ia y Ciencias, Universidad Diego Portales, Av. Ej\'ercito Libertador 441, Santiago, Chile}
\altaffiltext{6}{Department of Physics \& Astronomy, University of Leicester, Leicester, LE1 7RH, UK}
\altaffiltext{7}{Department of Space, Earth and Environment, Chalmers University of Technology, Onsala Space Observatory, Sweden}
\altaffiltext{8}{W. M. Keck Observatory, Waimea, HI 96743, USA}
\keywords{Galaxies: individual: WISEA J224607.56$-$052634.9 -- Galaxies: nuclei -- Galaxies: active -- Quasars: supermassive black holes -- Quasars: emission lines -- Infrared: galaxies }

\begin{abstract}
We use optical and near-­infrared spectroscopy to observe rest-­UV emission lines and estimate the black hole mass of WISEA J224607.56$-$052634.9 (\w2246m05) at $z = 4.601$, the most luminous hot dust-­obscured galaxy yet discovered by \WISE. From the broad component of the \MgII-­2799\AA\ emission line, we measure a black hole mass of $\log (M_{\rm BH}/M_{\sun}) = 9.6 \pm 0.4$. The broad \CIV-1549\AA\ line is asymmetric and significantly blue­shifted. The derived $M_{\rm BH}$ from the blueshift-­corrected broad \CIV\ line width agrees with the \MgII\ result. From direct measurement using a well-­sampled SED, the bolometric luminosity is $3.6\times 10^{14}\,L_{\sun}$. The corresponding Eddington ratio for \w2246m05\ is $\lambda_{\rm Edd} = L_{\rm AGN} / L_{\rm Edd} = 2.8$. This high Eddington ratio may reach the level where the luminosity is saturating due to photon trapping in the accretion flow, and be insensitive to the mass accretion rate. In this case, the $M_{\rm BH}$ growth rate in \w2246m05\ would exceed the apparent accretion rate derived from the observed luminosity.

\end{abstract}

\section{Introduction}

Discovered by their unusual mid-infrared colors in the \textit{Wide-field Infrared Survey Explorer} \citep[\WISE;][]{2010AJ....140.1868W} all-sky survey, Hot Dust-Obscured Galaxy \citep[Hot DOG;][]{2012ApJ...755..173E,2012ApJ...756...96W} are hyperluminous infrared galaxies with a wide IR plateau and a steep drop in the far IR, suggesting a broader dust temperature distribution \citep{2015ApJ...805...90T}, which we model in \cite{tsai_herschel}. \w2246m05 is the most luminous Hot DOG yet identified. With $L_{\rm bol} > 3 \times 10^{14} L_{\sun}$, it is well into the Extremely Luminous Infrared Galaxy \citep[ELIRG, $> 10^{14} L_{\sun}$;][]{2015ApJ...805...90T} range, and among the few most luminous galaxies known thus far.

Its superlative luminosity is not the result of gravitational lensing \citep{2015ApJ...805...90T,2016ApJ...816L...6D}. To account for a significant fraction of this luminosity via a starburst would require a star formation rate (SFR) $\gg 10^{4} M_{\sun}\,{\rm yr}^{-1}$ \citep{2015ApJ...805...90T}, but the cool gas and dust supplies needed for such an extraordinary SFR are not in evidence. Instead, the spectral energy distribution (SED) of \w2246m05\ is dominated by hot dust \citep[$T_{\rm dust} > 450$\,K;][]{2015ApJ...805...90T}, indicative of a dominant AGN. Direct evidence for an accreting supermassive black hole (SMBH) in this system comes from the broad \CIV\ line in its spectrum \citep{diaz-santos2017}, which we discuss in further detail below. This makes an obscured AGN the most straightforward power source for \w2246m05, and we assume this is the case for the remainder of this paper.

Like \w2246m05, many hyperluminous \hotdogs\ show AGN features in their spectra \citep{eisenhardt_spec}. Recently, \cite{Wu_bhm} detected broad \Ha\ lines in all members of a sample of five hyperluminous \hotdogs\ at $1.6 < z < 2.5$, finding black hole masses in the range $\log (M_{\rm BH}/M_{\sun}) =$ 8.7 -- 9.5. Compared to quasars with similar black hole masses, these \hotdogs\ have higher luminosities. The SMBHs in these \hotdogs\ are accreting at a rate close to the Eddington limit, suggesting that \hotdogs\ represent a transitional phase of high accretion between obscured and unobscured quasars \citep{Wu_bhm}. 

Is \w2246m05 similar to its sibling \hotdogs? Is its extreme luminosity due to sub-Eddington accretion onto an exceptionally massive SMBH, or to an exceptionally high Eddington ratio for a more typical SMBH mass \citep{2015ApJ...804...27A,2015ApJ...805...90T}? To answer these questions, the virial mass of the SMBH in \w2246m05\ needs to be determined. Measuring SMBH mass from \CIV\ profiles is challenging in comparison to Balmer lines, with large scatter \citep[e.g.][]{2007ApJ...671.1256N,2011ApJS..194...45S} and possible bias due to outflows \citep{1982ApJ...263...79G,1997ApJ...474...91M,2004ApJ...611..125L}. Although the \Ha\ and \Hb\ lines are stronger and suffer less from \FeII\ emission blending, for sources at higher redshift ($z > 4$), such as \w2246m05, only the broad \MgII\ line is reliable and available from the ground. 

In this paper, we report the detection of broad \MgII\ emission in \w2246m05. We provide black hole mass estimates from the \MgII\ and \CIV\ lines. To better determine the Eddington ratio, we also re-examine the luminosity estimate of \w2246m05\ with updated photometric data. We present our near-infrared spectroscopy of \w2246m05\ in Section \ref{sec:observations}, together with a description of other data used in this paper. Section \ref{sec:analysis} gives the SED, luminosity, and line widths based on these data. Section \ref{sec:results} considers the resulting black hole mass and Eddington ratio. In Section \ref{sec:conclusions}, we summarize our work. A redshift $z = 4.593$ for \w2246m05\ was reported by \cite{2015ApJ...805...90T}, determined from the overall Ly$\alpha$ line profile. However, the redshift was revised to $z = 4.601$ based on the \CII\,157.7\micron\ line emission \citep{2016ApJ...816L...6D}. We adopt $z = 4.601$ for \w2246m05, and a cosmology with $H_{0} = 70$ km\,s$^{-1}$\,Mpc$^{-1}$, $\Omega_{m} = 0.3$, and $\Omega_{\Lambda} = 0.7$.

\section{Observations and Data Reduction}\label{sec:observations}

\begin{figure}
\epsscale{1.1}
\begin{center}
\plotone{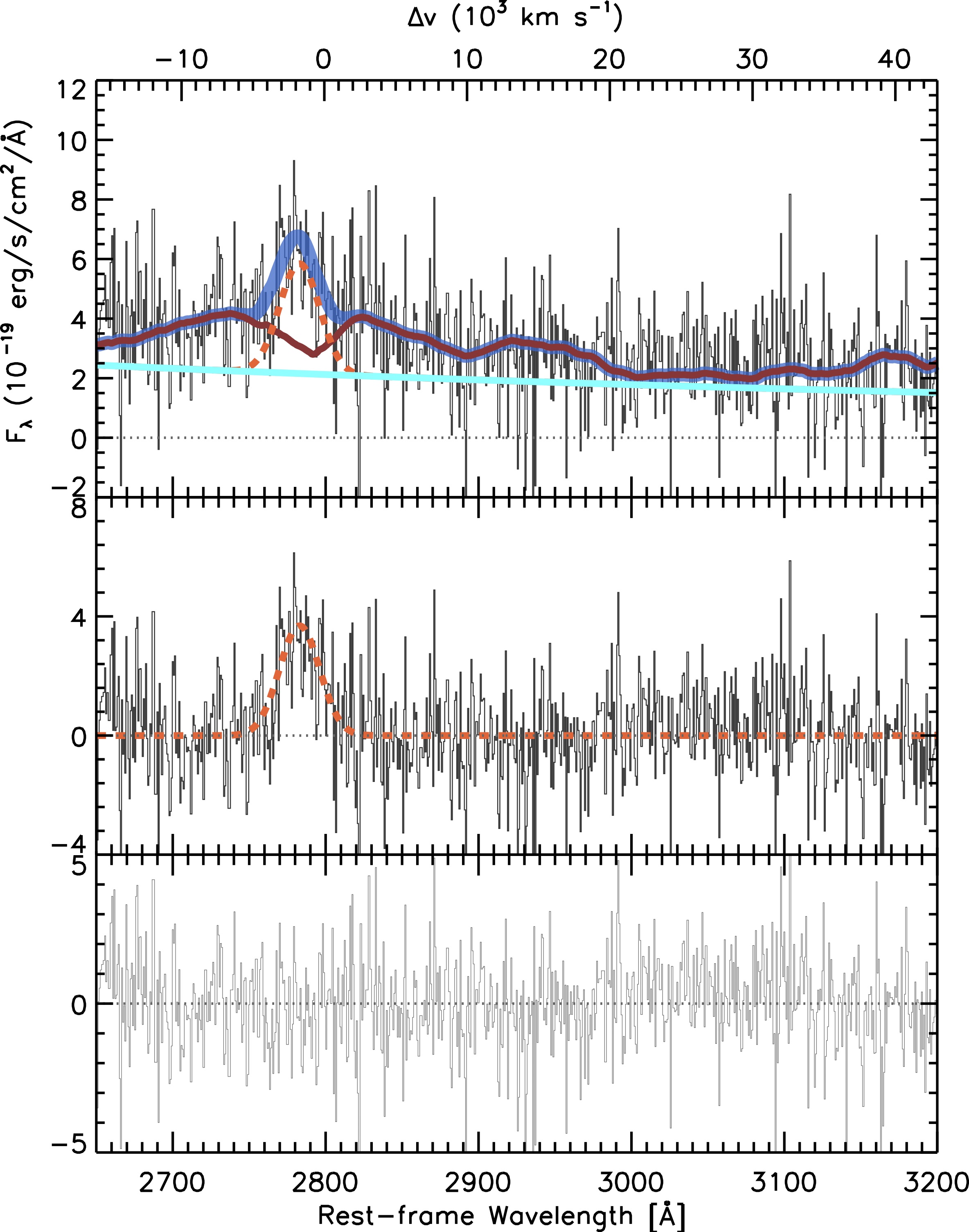}
\end{center}
\caption{
The observed near-IR spectrum of \w2246m05\ around the \MgII\ 2799\AA\ emission line, with a model (blue solid line) containing a Gaussian profile for the \MgII\ line (orange dashed line), the \FeII\ line complex (brown solid line), and a power-law continuum ($F_{\lambda}$ [$10^{-19}$\,erg\,s$^{-1}$\,cm$^{2}$ \AA$^{-1}$] = $1.808  \times (\lambda / 3000 \AA)^{-2.583} $; cyan solid line). Velocities are shown with respect to the \CII\ redshift $z = 4.601$. The middle panel shows the spectrum and \MgII\ line component (FWHM $= 3300\pm600$\,\kms, blueshifted by $1600\pm300$\,\kms) after the continuum and \FeII\ complex are subtracted. The residual spectrum is plotted in the lower panel. 
}\label{fig:MgII}
\vspace{0.2mm}
\end{figure}

\subsection{Keck OSIRIS Observation of the \MgII\ Line}

At $z = 4.601$, the \MgII-2799\AA\ line of \w2246m05\ falls at 1.567\,\micron\ in the $H$ band. \w2246m05\ was observed with Keck I telescope on UT 2016 October 21 in the first half of the night using OSIRIS \citep{2006SPIE.6269E..42L} while the originally proposed instrument (MOSFIRE) was being repaired. The weather was clear and the seeing was $\sim 0\farcs7$ over the observation. The spectra were taken with the \textit{Hbb} filter covering 1.473 -- 1.803 \micron\ at 0.5 nm channel$^{-1}$ resolution. With the $4 \times 4$ dithering used, the 1\farcs6 $\times$ 6\farcs4 field-of-view of the integrated field unit covered $2\farcs0 \times 10\farcs9$ with $0\farcs1$ pixel$^{-1}$ at a position angle of 337\degr\ centered on \w2246m05.  The final spectral resolution of $\sim 2.5$ spaxels yielded $R = \lambda/\Delta\lambda \lesssim 3400$. Atmospheric wavefront errors were partially corrected with Laser Guild Star Adaptive Optics (LGS-AO) and a $R = 18.3$ tip-tilt star 26\farcs8 to the south-west of \w2246m05. The integration time for each individual frame was 900 s, and a total of 4 hours (16 frames) of on-source data was collected. 

The OSIRIS data were reduced with the OSIRIS data reduction pipeline (DRP) v4.0.0. A custom procedure was used to remove the sky emission using pixels near the science target and frames adjacent in time. About 6\%\ of the data at wavelengths with substantial sky line residual presents were clipped. The telluric correction was applied using both the G2V star HD 216516, and the A0V star HD 219833. The flux calibration was done using the G2V star. Including the uncertainties due to calibrator flux, aperture correction, and Strehl ratio, the overall uncertainty in the calibrated fluxes is estimated to be $\sim$ 20\%. The wavelength was calibrated to values in vacuum. The final mosaic data cube was produced using the LGS offsets between frames. Using the 2MASS point source catalog \citep{2006AJ....131.1163S} and the WFC3 image in F160W from \textit{HST} \citep{2016ApJ...816L...6D}, we registered the final science data cube with the uncertainty in absolute astrometry estimated to be $<$ 0\farcs1. 

Due to the LGS-AO, the spatial resolution was sufficient to resolve the continuum profile of \w2246m05, with an estimated extent of $0\farcs36 \times 0\farcs29$ at PA$=155\degr$, consistent with the \textit{HST} observations \citep{2016ApJ...816L...6D}. The \MgII\ emission line of \w2246m05\ was extracted from the OSIRIS data cube using a 0\farcs3 aperture. An aperture correction was applied to the flux scale of the \MgII\ spectrum to match the photometry from the \HST\ F160W image. The extracted and flux corrected spectrum is shown in Figure \ref{fig:MgII}.

\begin{deluxetable}{lrrlc}[t] 
\tabletypesize{\scriptsize}
\tablewidth{0in}
\tablecaption{Fluxes of \w2246m05 \label{table:fluxes}}
\tablehead{
\multicolumn{1}{c}{Band} &
\multicolumn{1}{c}{Wavelength} &
\multicolumn{1}{c}{Flux Density} &
\multicolumn{1}{c}{Reference}}
\startdata
WFC3 $F160W$		&	1.537 \micron	&	$6.1\pm0.2$\,$\mu$Jy	&	\textbf{\textit{D16}}	\\
$K$-band					&	2.159 \micron	&	$8.9\pm2.8$\,$\mu$Jy	&	\textbf{\textit{A15}}	\\
IRAC band 1			&	3.6 \micron	&	$38\pm2$ $\mu$Jy		&	\textbf{\textit{G12}}	\\
IRAC band 2			&	4.5 \micron	&	$33\pm1$ $\mu$Jy		&	\textbf{\textit{G12}}	\\
$\WISE$ band 3			&	12 \micron		&	$2.5\pm0.2$\,mJy	&	\textbf{\textit{T15}}	\\
$\WISE$ band 4			&	22 \micron		&	$15.9\pm1.6$\,mJy	&	\textbf{\textit{T15}}	\\
PACS blue channel			&	70 \micron		&	$37\pm3$\,mJy	&	\textbf{\textit{T15}}	\\
PACS red channel			&	160 \micron	&	$142\pm16$\,mJy	&	\textbf{\textit{T15}}\tablenotemark{a}	\\
SPIRE 250 \micron			&	250 \micron	&	$107\pm8$\,mJy	&	\textbf{\textit{T15}}\tablenotemark{a}	\\
SPIRE 350 \micron			&	350 \micron	&	$81\pm12$\,mJy	&	\textbf{\textit{T15}}	\\
SCUBA2 450 \micron		&	450 \micron	&	$49\pm12$\,mJy		&	\textbf{\textit{J14,T18}}	 \\
SPIRE 500 \micron			&	500 \micron	&	$44\pm15$\,mJy	&	\textbf{\textit{T15}}	\\
SCUBA2 850 \micron		&	850 \micron	&	$11\pm2$\,mJy		&	\textbf{\textit{J14}} \\
ALMA 882 \micron			& 	882 \micron	& 	$7.4\pm0.4$\,mJy	&	\textbf{\textit{D16}}\tablenotemark{b} \\
ALMA 1.2 mm				&	1190 \micron	&	$4.8\pm1.9$\,mJy	&	\textbf{\textit{D18}}\tablenotemark{c} 
\enddata
\tablecomments{The reference code in the last column: \textbf{\textit{A15}}: \cite{2015ApJ...804...27A}; \textbf{\textit{G12}}: \cite{2012AJ....144..148G}; \textbf{\textit{D16}}: \cite{2016ApJ...816L...6D}; \textbf{\textit{D18}}: \cite{diaz-santos2017}; \textbf{\textit{J14}}: \cite{2014MNRAS.443..146J}; \textbf{\textit{T15}}: \cite{2015ApJ...805...90T}; \textbf{\textit{T18}}: This work. $^{\rm a}$\,The \Herschel\ PACS 160 \micron\ and SPIRE 250 \micron\ measurements are updated. See text for details. $^{\rm b}$\,Flux within 1\arcsec\ diameter aperture. $^{\rm c}$\,Sum of \w2246m05\ and the extended emission.  
}
\end{deluxetable}  


\begin{figure*}
\begin{center}
\epsscale{1}
\plotone{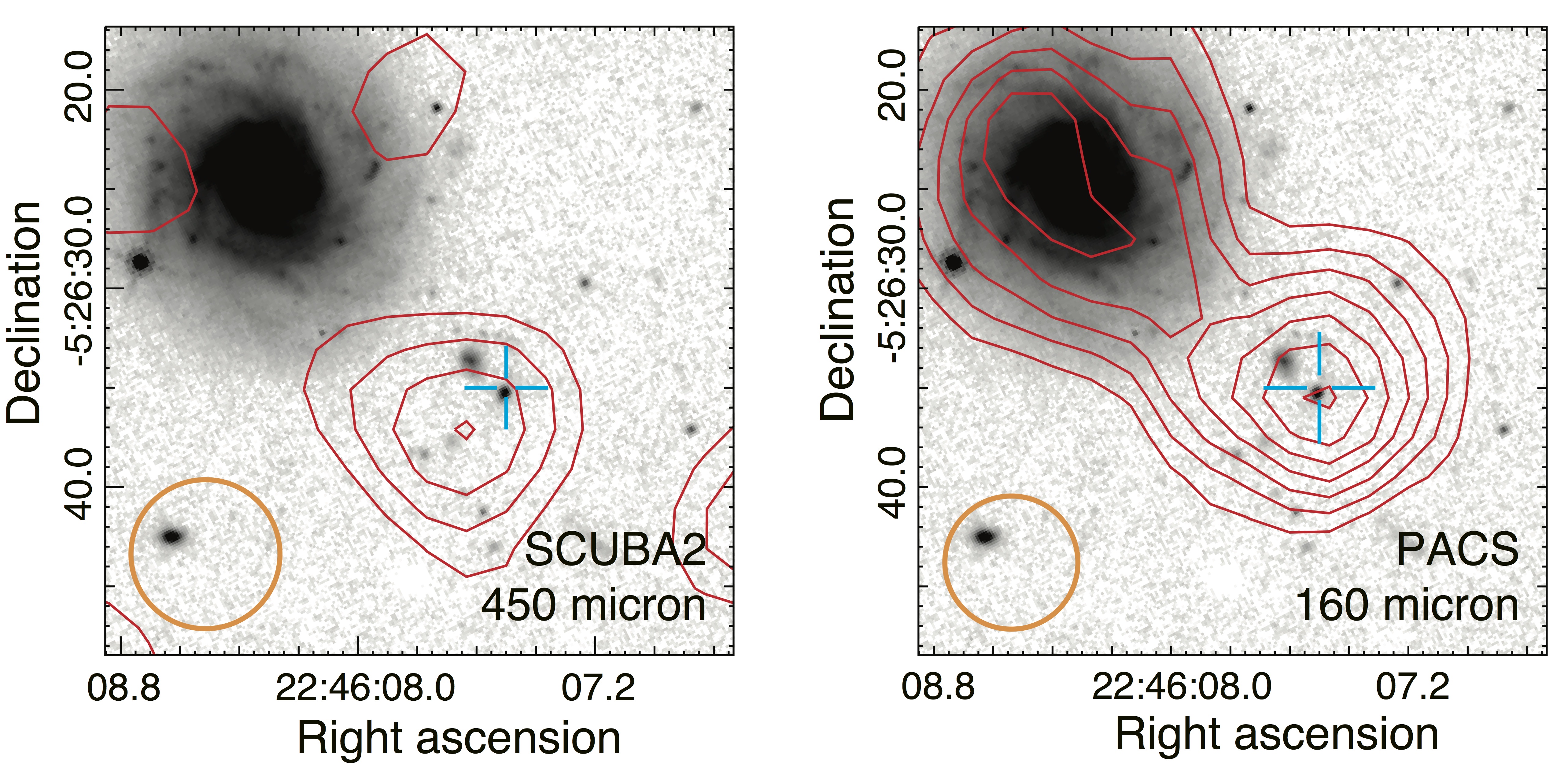}
\end{center}
\caption{
(\textbf{\textit{Left}}) SCUBA2 450\,\micron\ contours at levels with a increment of 12 mJy\,beam$^{-1}$ ($1\,\sigma$) and (\textbf{\textit{Right}}) \herschel\ PACS 160\,\micron\ contours at levels of 30\%\ -- 100\%\ by 10\% steps overlaid on the \HST\ F160W image of \w2246m05. The \wise\ position of the source is marked with the blue crosshair. The FWHM of the PACS 160\,\micron\ PSF is 6\farcs3 and the beam size at 450\,\micron\ is 7\farcs5, as shown at the lower right corners of the corresponding panels (orange). The peak of the 450\,\micron\ emission is marginally offset to the south-east of the \wise\ position by 2\farcs4 $\pm 2\farcs2$.  The foreground galaxy to the north-east at $z_{\rm phot} = 0.047$ shows no significant detection at  450\,\micron, and is unlikely to significantly contaminate \herschel\ photometry at 250\,\micron -500\,\micron.  
}\label{fig:scuba2_450um_PACS_160um}
\end{figure*}

\subsection{Other Data}

For comparison to the SMBH mass estimate from \MgII\, we analyze an optical spectrum of the \CIV\ line.  To calculate the Eddington ratio, we estimate the luminosity using the observed spectral energy distribution (SED) of \w2246m05\ from optical to submillimeter wavelengths. The flux densities and the references are listed in Table \ref{table:fluxes}. These observations are described below.

\subsubsection{Keck LRIS Observation of the \CIV\ Line}

As reported in \cite{diaz-santos2017}, a one hour exposure optical spectrum of \w2246m05\ was obtained on UT 2013 October 5 using LRIS \citep{1995PASP..107..375O} on the Keck I telescope, with spectral resolution $\sim 750$ and a 1\farcs5 slit. Additional details of the observation are reported in \cite{diaz-santos2017}. 

\subsubsection{SCUBA2 450\,\micron\ Observation}

SCUBA2 observations with the JCMT, as reported by \cite{2014MNRAS.443..146J}, were obtained on UT 2012 May 23 and 26. During the 450\,\micron\ observations, the CSO 225\,GHz sky opacity was $\tau_{225} \sim 0.05$, and the corresponding optical depth at 450\,\micron\ was $\tau_{\rm 450\micron} \sim 1$. However, the two sets of observations identify a point source with a consistent flux density (within $1\sigma$), slightly offset to the south­east from the \wise\ ­infrared position of \w2246m05. The position offset ($2\farcs4 \pm 2\farcs2$) is consistent with the pointing uncertainty during the observation. The total time per source was 120 min using the CV DAISY mode, providing deep coverage in the central 3\arcmin\ diameter region. The final map that combines both sets of observations is shown in the left panel of Figure \ref{fig:scuba2_450um_PACS_160um}. 

\subsubsection{\Herschel\ Observations}

Herschel fluxes for \w2246m05\ were reported in \cite{2015ApJ...805...90T}. There is a foreground spiral galaxy \citep[SDSS J224608.38$-$052624.3;][photometric redshift 0.047$\pm$0.024]{2003AJ....125..580C,2016MNRAS.460.1371B} $\sim$ 16\farcs5 to the north-east of \w2246m05, with $r_{\rm \sc PETRO} = 8\farcs3$ (Figure \ref{fig:scuba2_450um_PACS_160um}). With the $\sim$ 12\arcsec\ FWHM beam size for PACS at 160 \micron\, this could affect the \Herschel\ fluxes of \w2246m05\ reported in \citet{2015ApJ...805...90T}.  For the SPIRE bands, the larger beam size (18\arcsec\ to 37\arcsec) increases the possibility of flux contamination. This concern has been raised by \cite{2018ApJ...854..157F}, who have tried to estimate the contribution of the foreground galaxy to the \Herschel\ photometry with PSF fitting and SED modeling. However, the estimated flux contamination in that work does not reconcile with the results of the \Herschel/PACS Point Source Catalogue \citep{2017arXiv170505693M}, in which both the foreground galaxy and \w2246m05\ are detected using PSF photometry. We remeasured the 160\,\micron\ flux density of \w2246m05\, excluding emission from the extended profile of the foreground galaxy by using an irregular polygon aperture, finding $F_{\rm 160\micron} \sim$ 142 mJy, 25\% less than reported in \citet{2015ApJ...805...90T}. All other \Herschel\ photometry in \citet{2015ApJ...805...90T} agrees with the results from the \Herschel/PACS Point Source Catalogue \citep{2017arXiv170505693M} and \Herschel/SPIRE Point Source Catalogue \citep[SPSC;][]{2017arXiv170600448S} within $1 \sigma$, except the flux density for SPIRE at 250\micron\ which is $\sim 2\sigma$ lower than the SPSC value. For consistency, we adopt the SPSC flux at 250\micron. We note that the SPIRE maps all show a point source with the peak within the corresponding FWHM from \w2246m05. Our ground-based JCMT SCUBA2 maps at 450\,\micron\ and 850\,\micron, with beam sizes of 7\farcs5 and 15\arcsec, also show that the observed submillimeter emission is concentrated at \w2246m0526\ \citep{2014MNRAS.443..146J} with no significant emission detected around the foreground galaxy (see Figure \ref{fig:scuba2_450um_PACS_160um}--\textit{left}). \w2246m05\ dominates the observed far-IR and submillimeter emission over the adjacent spiral galaxy. Using the peak flux at the pixel of the foreground galaxy center, we estimate the SPIRE band photometry contamination by the foreground source is $<$ 30\%. From the ALMA 252 GHz continuum map with a 20\farcs2 half-power-beam-width primary beam \citep{diaz-santos2017}, the dust emission from the foreground galaxy is $<$ 10\% of the flux of \w2246m05\ system at 1.2\,mm. 

\section{Analysis}\label{sec:analysis}

\begin{figure*}
\epsscale{0.8}
\begin{center}
\plotone{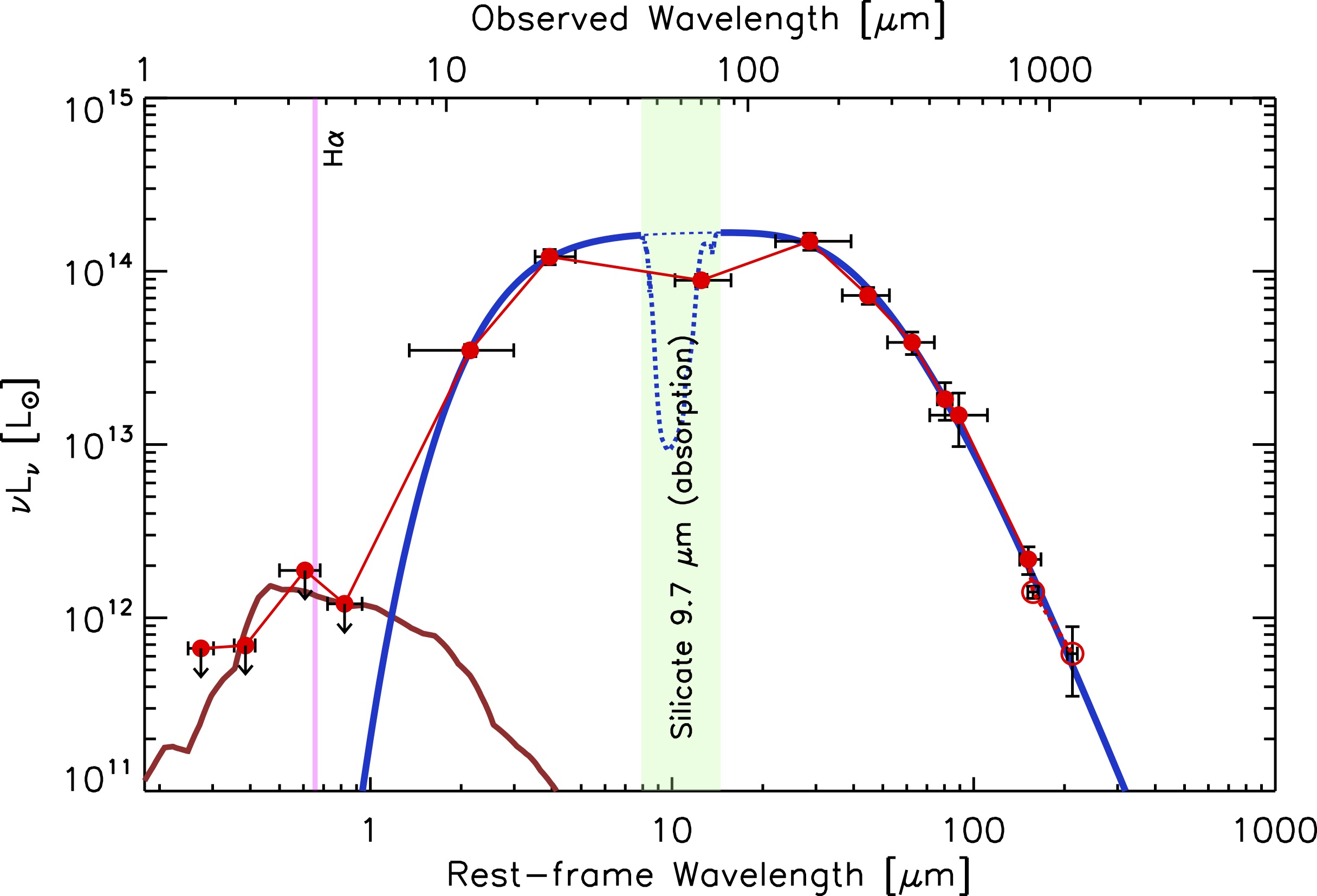}
\end{center}
\caption{
SED of \w2246m05. The filled red points represent photometric measurements with detections ($> 3 \sigma$). The two open red circles are the integrated flux from ALMA measurements of the resolved emission. The downward arrows at rest-frame wavelengths shorter than 1 \micron\ indicate that these data are only used as upper limits in the SED fitting. The light green shaded area indicates the wavelength range of a broad silicate absorption feature which could be the cause of the PACS 70\,\micron\ flux deficit. The blue line shows an SED model with a continuous dust temperature distribution \citep{tsai_herschel} with an empirical silicate absorption feature template scaled to match the photometry (dotted blue line). The magenta line marks the wavelength of the \Ha\ line. The brown solid line represents the SED of a 1.3 Gyr-old elliptical galaxy with a total stellar mass of $1.1 \times 10^{12} M_{\sun}$ generated with the GRASIL code.
}\label{fig:SED}
\end{figure*}

\subsection{SED and Luminosity}

The SED of \w2246m05\ is shown in Figure \ref{fig:SED}, based on the updated photometric data listed in Table \ref{table:fluxes}. Unlike optical QSOs in which a large fraction of energy escapes at UV wavelengths, most of the energy from \w2246m05 is seen at rest-frame infrared wavelengths ($> 1$\,\micron). The broad \MgII\ and \CIV\ emission lines observed in \w2246m05 provide direct evidence for the presence of an AGN, and the infrared emission most plausibly arises from hot dust obscuring the AGN.  The rest-frame UV and optical continuum emission, which are assumed to be primarily from the host galaxy of the obscured AGN, contribute $\lesssim$ 1\%\ of the total luminosity. Like other \hotdogs, the infrared SED of \w2246m05\ shows this plateau, but, interestingly, also shows a dip with respect to other \hotdogs\ at rest-frame $\sim 12.5$\,\micron\ (70\,\micron\ in the observed frame). This prompts us to suggest that silicate absorption may be affecting the SED.

At $z=4.601$, the broad 9.7\,\micron\ silicate absorption feature overlaps with the 60 - 85\,\micron\ bandpass of the PACS 70\,\micron\ filter, and can account for the dip in the SED if the strength of the absorption in \w2246m05\ is comparable to that in the heavily enshrouded nucleus of NGC\,4418 \citep{1986MNRAS.218P..19R,2001A&A...365L.353S}, the ULIRG Arp\,220 \citep{2007ApJ...663...81P}, or the HyLIRG IRAS\,08572$+$3915 \citep[][]{2007ApJ...654L..49S,2008A&A...484..631V,2014MNRAS.437L..16E}. In Figure \ref{fig:SED}, we plot a template of the silicate absorption in NGC\,4418 scaled to match the observed 70\,\micron\ photometry (for clarity we omit showing the relatively weak 9.66 \micron\ \Htwo\ 0--0 S(3) emission line).

We find the bolometric luminosity of \w2246m05\ is $L_{\rm bol} = 3.6\pm0.3 \times 10^{14}\,L_{\sun}$,  using the observed flux density measurements listed in Table \ref{table:fluxes}. This estimate follows the methodology of \cite{2015ApJ...805...90T} by integrating a power law interpolated between photometric data. We assume essentially all of this luminosity comes from a quasar shrouded within a dust cocoon.

Hot dust emission dominates the energy output, as indicated by the SED. Because of the high luminosity, the dust sublimation radius is $\sim 15$\,pc for \w2246m05\ \citep{1987ApJ...320..537B,2015ApJ...805...90T}, substantially larger than the radius of the broad line region \citep[$\sim$ 1.3\,pc, based on][]{2009ApJ...697..160B}. Thus the thermal dust emission can not vary dramatically within the light-crossing rest-frame timescale of $\sim 50$ yr, or $\sim 280$ yr in the observed frame. This timescale is even longer for the dust emission at the longer wavelengths of the SED plateau. Therefore we do not anticipate observable variability in the bolometric luminosity of \w2246m05.

\subsection{\MgII\ Emission Line Width}\label{sec:MgII_line_width}

The observed \MgII\ 2799\AA\ emission line and the line model are shown in Figure \ref{fig:MgII}, which covers the spectrum at rest-frame wavelengths between 2630\AA--3220\AA. The signal-to-noise ratio of the continuum and the \MgII\ line (within FWHM from the line center) are 1.4 and 7.7 per spectral element ($0.82\,\AA$, rest), respectively.

The \MgII\ doublet line profile was fit after modeling the blended \FeII\ line complex and a power-law continuum. The profile of the \FeII\ complex used templates from \citep{2006ApJ...650...57T}. Least-squares model fitting was done using the Levenberg-Marquardt algorithm as implemented in IDL MPFIT \citep{2009ASPC..411..251M}. Excluding the \MgII\ region, we stepped through a grid of \FeII\ widths with FWHM=0 to 18000 \kms\ and a 600 \kms\ step size, finding the best fit continuum (given in the Figure \ref{fig:MgII} caption) and \FeII\ model strength. We then fit the residual with a single Gaussian model for the \MgII\ line, iterating up to 20 times until the parameters stay unchanged to floating point precision.

Although the signal-to-noise ratios are not high, they do not significantly affect the reliability for the line profile measurements, because of the broad line width. The best-fit model yields a Gaussian with a FWHM of $3300\pm600$\,\kms, and a blueshifted velocity offset of $\Delta v = 1600\pm300$\,\kms\ with respect to the \CII\ redshift \citep{2016ApJ...816L...6D}.

\subsection{\CIV\ Emission Line Width}\label{sec:CIV_line_width}

\begin{figure}
\epsscale{1.1}
\begin{center}
\plotone{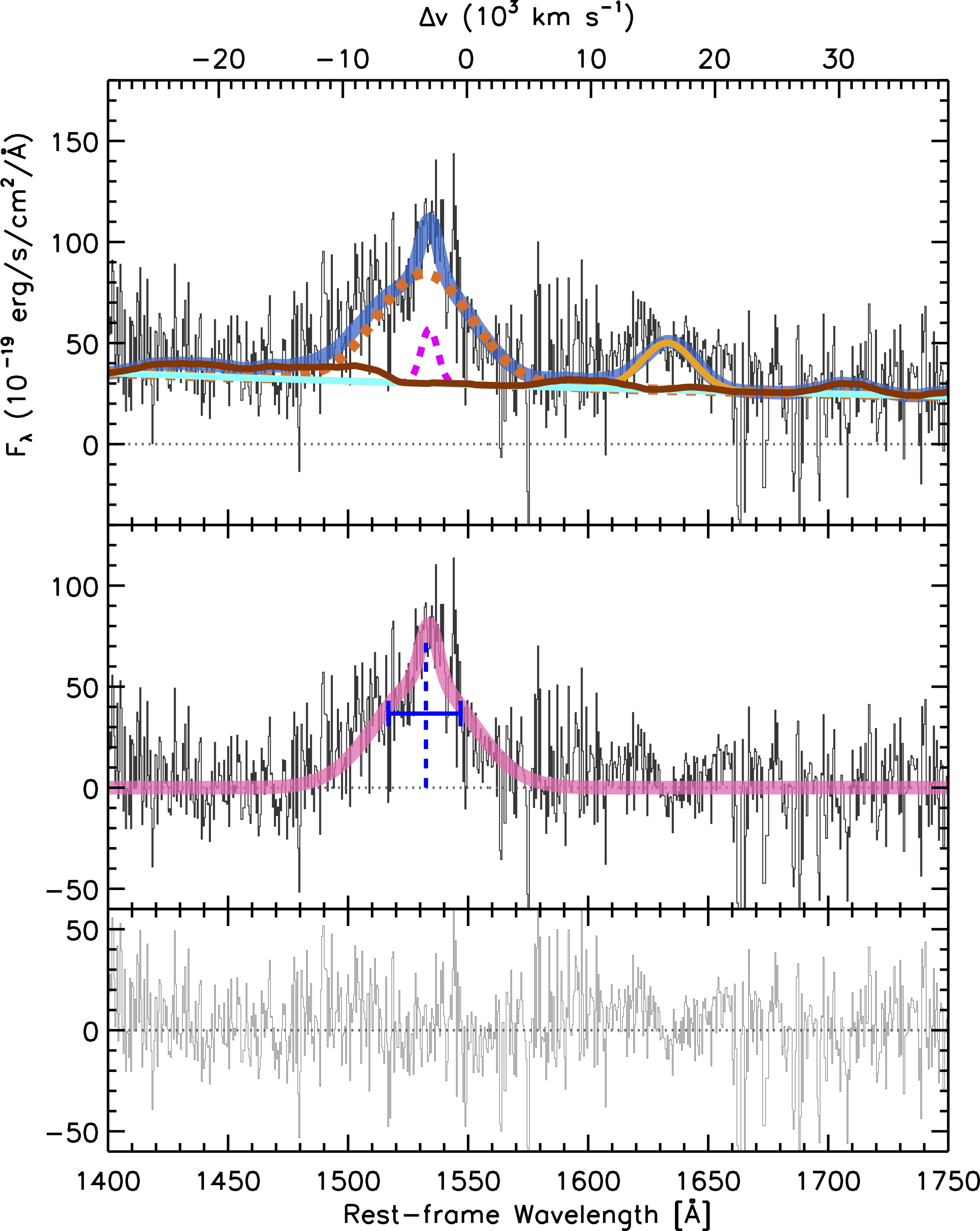}
\end{center}
\caption{
The observed optical (rest UV) spectrum of \w2246m05\ is shown in the top panel, with a model (blue solid line) containing two independent Gaussian \CIV-1488,1551\AA\ emission line profiles (orange and magenta dashed lines), the \HeII-1640\AA\ line (orange solid line), the \FeII\ line complex (brown solid line), and a power-law continuum ($F_{\lambda}$ [$10^{-19}$\,erg\,s$^{-1}$\,cm$^{2}$ \AA$^{-1}$] = $38.00 \times (\lambda / 1350 \AA)^{-1.784} $; cyan solid line). Velocities are shown with respect to the \CII\ redshift of $z = 4.601$. The middle panel shows the composite of the two Gaussian models (FWHM $= 5900\pm100$\,\kms, blueshifted by $3300\pm70$\,\kms) overlaid on the observed spectrum after subtracting the other model components. The residual spectrum is presented in the lower panel.
}\label{fig:CIV}
\end{figure}

At $z = 4.601$, the \CIV\ line shifts to $\sim 8674$\,\AA. As noted in \cite{diaz-santos2017}, the \CIV\ line in \w2246m05 is highly asymmetric, broad, and significantly blue-shifted relative to this ALMA-derived \CII\ redshift.  Following a similar approach to that used by \cite{2008ApJ...680..169S,2011ApJS..194...45S} and \cite{2017ApJ...838...41J}, we model the \CIV\ line of \w2246m05\ with two components as shown in Figure \ref{fig:CIV}. The signal-to-noise ratio of the continuum and the \CIV\ line are 4.2 and 13.4 per spectral element ($2.1\,\AA$, rest), respectively.

As we did for \MgII, we first solved for a power-law continuum (given in the Figure \ref{fig:CIV} caption) and the \FeII\ line complex \cite[using the template from][]{2001ApJS..134....1V}, as well as the \HeII-1640\AA\ line. The strength of the \FeII\ complex was matched to that of the \FeII\ around the \MgII\ region using overlapping wavelengths in the templates, and the \FeII\ line widths were set to be identical for the \MgII\ and \CIV\ regions.

The narrower Gaussian component has a FWHM $= 1600\pm140$\,\kms, blueshifted by $3000\pm80$\,\kms\ with respect to the system redshift, while the broader component has FWHM $= 9000\pm140$\,\kms\ and is blueshifted by $3420\pm70$\,\kms. The composite double-Gaussian profile has FWHM $= 5900\pm100$\,\kms\ and a blueshift of $3300\pm70$\,\kms. This blueshift of the \CIV\ line profile in \w2246m05\ is significantly higher than the median blueshift observed in SDSS QSOs \citep[median = 890\,\kms\ for 492 quasars at $4.5 < z < 4.7$;][]{2011ApJS..194...45S}, and more comparable to the median blueshift of 2520\,\kms\ for 9 quasars at $z \gtrsim 6.4$ that have \CIV\ emission line measurements \citep{2017ApJ...849...91M}.

\section{Results and Discussion}\label{sec:results}

\subsection{\MgII-based SMBH Mass}\label{sec:MgII_MBH}

Masses of the SMBH in AGNs are usually determined from the measurements of broad emission lines, assuming virialized gas motion. This approach was established via variability monitoring \citep[reverberation mapping; see e.g.][]{1993PASP..105..247P} and subsequently calibrated to single epoch measurements of the H$\beta$ line width and 5100\,\AA\ continuum \citep[e.g.][]{1999ApJ...526..579W,2000ApJ...533..631K,2002ApJ...571..733V}, and extended to other line widths.  For objects at $z > 1$, the H$\beta$ line is redshifted into the infrared, stimulating the use of the broad \MgII\ line from optical spectroscopy for SMBH mass determination \citep{2002MNRAS.337..109M,2008ApJ...689L..13O,2012MNRAS.427.3081T,2013A&A...555A..89M}. Advances in near-IR instrumentation have made near-IR spectroscopy of distant objects feasible, enabling the Balmer lines to be used for SMBH measurement to $z \sim$ 2.5--3.5 \citep[e.g.][]{Wu_bhm}. At $z > 3.5$ the \MgII\ line is also observed at $> 1\mu$m, and \MgII\ is usually used for black hole mass ($M_{\rm BH}$) measurement of objects at the highest redshifts \citep{2015Natur.518..512W,2018Natur.553..473B}. The calibration of \MgII-determined SMBH masses with respect to those from \Hb\ has been established \citep{2004MNRAS.352.1390M,2009ApJ...707.1334W,2011ApJS..194...45S,2015ApJ...806..109J,2015ApJ...815..129S}. In this paper, we adopt the \MgII-based SMBH mass formulation from Equation 10 of \cite{2009ApJ...707.1334W}:
\begin{eqnarray}\label{eq:Mbh_MgII}
\log \left( \frac{M_{\rm BH}}{M_{\sun}} \right) =&& (7.13\pm0.27) + 0.5\,\log \left( \frac{L_{3000}}{10^{44}\,{\rm erg\,s^{-1}}} \right) \nonumber\\
&& +\, (1.51\pm0.49)\,\log \left( \frac{\rm FWHM_{Mg\,II}}{1000\,{\rm km\,s^{-1}}} \right), 
\end{eqnarray}
where $L_{3000}$ is the monochromatic luminosity at rest-frame 3000\AA, and ${\rm FWHM_{Mg\,II}}$ is the full-width-half-maximum of the \MgII\ line profile (given in section \ref{sec:MgII_line_width}).

Because of the high extinction in \w2246m05, estimates of $L_{3000}$ from the rest-UV continuum are uncertain. Instead, we use $L_{3000} \sim 0.19 \times L_{\rm bol}$ based on the empirical unobscured AGN SED model of \citet{2006ApJS..166..470R}. For $L_{\rm bol} = 3.6 \times 10^{14}\,L_{\sun}$, $L_{3000} = 2.6 \times 10^{47}\,{\rm erg\,s^{-1}}$. The observed rest-frame 3000\AA\ continuum flux in \w2246m05 is only about 0.02 of the anticipated value from this template. With the \MgII\ FWHM of $3300\pm600$\,\kms, this yields $\log (M_{\rm BH}/M_{\sun}) = 9.6 \pm 0.4$. The error range includes the $1\,\sigma$ systematic uncertainties from Equation \ref{eq:Mbh_MgII} and from the \MgII\ profile fitting. The statistical uncertainty of $M_{\rm BH}$ due to the \MgII\ line width uncertainty is 0.05 dex.

The \citet{2006ApJS..166..470R} SED model assumes the emission from a quasar is isotropic. However, the unobscured quasars used to construct the SED presumably have a surrounding dusty torus with a low inclination and a covering factor that intercepts and reprocesses some of the quasar emission, so that it appears again in the IR. Thus the $L_{\rm bol}$ from the SED model would be overestimated by $(1 + CF_{\rm u})$ where $CF_{\rm u}$ is the covering factor for the unobscured quasars used in \citet{2006ApJS..166..470R}, and $L_{3000}$ would represent a larger portion of the true $L_{\rm bol}$. For \w2246m05\ which has a well determined $L_{\rm bol}$, this suggests using a ratio of $L_{3000}$/$L_{\rm bol}$ which is $(1 + CF_{\rm u})$ larger. The covering factor cannot be higher than 1 (especially for an unobscured quasar sample), so using Equation \ref{eq:Mbh_MgII}, this correction would increase $\log (M_{\rm BH}/M_{\sun})$ by $< 0.15$. Both of these terms are smaller than other estimates of the overall systematic uncertainty, which are up to 0.3 dex \citep{2009ApJ...692..246D,2009ApJ...707.1334W,2015ApJ...806..109J}.

\subsection{\CIV-based $M_{\rm BH}$ Estimate}\label{sec:CIV_MBH}

The \CIV\ line profiles of AGNs often show an enhanced blue wing, significantly different from their \Hb\ line profiles. This highlights the issue of the virial assumption for \CIV, and results in a large and biased offset when comparing \CIV-based $M_{\rm BH}$ estimates to those based on \Hb\ and \MgII\ \citep{2007ApJ...671.1256N,2011ApJS..194...45S}. In addition, the broad \CIV-1549\AA\ feature is often substantially blueshifted with respect to the system rest frame \citep[e.g.][]{2002AJ....124....1R}, especially for high luminosity objects \citep{2005MNRAS.356.1029B}. The blueshift is usually attributed to the wind component of the broad line region, or outflow \citep{1982ApJ...263...79G,1997ApJ...474...91M,2004ApJ...611..125L}.  Nevertheless, because of the accessibility of the \CIV\ line from the ground for the quasars over a large range of redshift ($1.3 \lesssim z \lesssim 5$), efforts have been made to calibrate \CIV-based $M_{\rm BH}$ estimates to \Hb-based values in Type-1 AGNs \citep{2006ApJ...641..689V,2011ApJ...742...93A,2011ApJS..194...45S,2012ApJ...759...44D,2013ApJ...775...60D,2013MNRAS.434..848R,2017ApJ...839...93P} although \CIV\ is not considered as reliable as \MgII\ \citep{2010MNRAS.409..591F,2015ApJ...806..109J,2016MNRAS.460..187M}. Motivated by the significance of the \CIV\ asymmetry in a principal component analysis of AGNs \citep{2007ApJ...666..757S}, a simple correction to the \CIV-based $M_{\rm BH}$ estimate has recently been suggested \citep{2015MNRAS.451.1290B,2016MNRAS.461..647C,2017MNRAS.465.2120C,2017ApJ...838...41J}. This correction calibrates the FWHM of the \CIV\ line to the expected virial \CIV\ FWHM from the Balmer line widths using the \CIV\ blueshift. 

As noted in section \ref{sec:CIV_line_width}, the \CIV\ line in \w2246m05 is highly asymmetric, broad, and significantly blue-shifted relative to the [CII] redshift. Using the composite Gaussian model and adopting the correction of \cite{2017MNRAS.465.2120C}:
\begin{eqnarray}
{\rm FWHM}&&_{\rm C\,IV}^{\rm Corr.} \nonumber\\
=&& \frac{\rm FWHM_{\rm C\,IV}^{\rm Measured}}{(0.36\pm0.03)\left( \frac{{\rm C\,IV}_{\rm blueshift}}{\rm 10^{3}\, km\,s^{-1}} \right) + (0.61\pm0.04)},
\end{eqnarray}
we obtain a corrected FWHM for the broad \CIV\ component of ${\rm FWHM}_{\rm C\,IV}^{\rm Corr.} = 3300\pm400$\,\kms, a factor 1.8 smaller than the FWHM derived from our two component analysis. 

To estimate the black hole mass, we adopt the methodology of \cite{2017MNRAS.465.2120C}:
\begin{eqnarray}
\log \left(  \frac{M_{\rm BH}}{M_{\sun}} \right) = && 6.71 + 2\,\log \left( \frac{\rm FWHM_{\rm C\,IV}^{\rm Corr.}}{\rm 10^{3}\, km\,s^{-1}} \right) \nonumber\\
&& +\, 0.53\, \log \left( \frac{L_{1350}}{\rm 10^{44}\, erg\,s^{-1}} \right).
\end{eqnarray}
$L_{1350}$ is the monochromatic luminosity at 1350\AA, and is estimated to be $L_{1350} \sim 0.26 \times L_{\rm bol} = 3.6\times 10^{47}\,{\rm erg\,s^{-1}}$ using the AGN template of \cite{2006ApJS..166..470R}.

Using the corrected value for the broad \CIV\ component, this yields $\log (M_{\rm BH}/M_{\sun}) = 9.6\pm0.4$ including systematic uncertainty, nearly identical to the $M_{\rm BH}$ estimated using the \MgII\ line. The agreement of the $M_{\rm BH}$ estimate using the \CIV\ line profile with that from the \MgII\ line measurement may be coincidental. It has been argued that a large component of the broad \CIV\ emission line is observed to not reverberate for nearby AGNs based on reverberation mapping studies \citep[e.g.,][]{2012ApJ...759...44D}. 

\begin{figure*}
\epsscale{0.75}
\begin{center}
\plotone{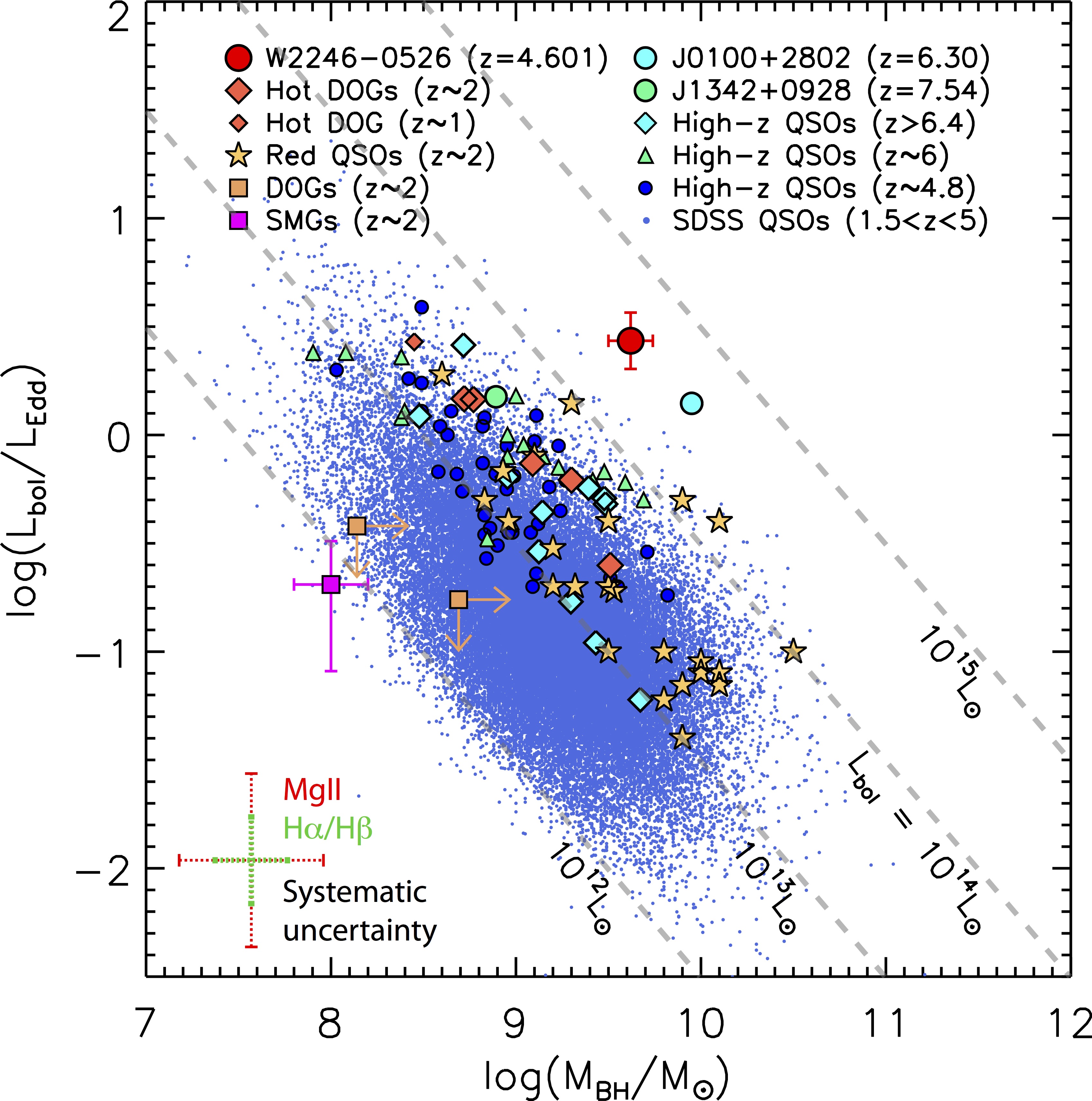}
\end{center}
\caption{
Eddington ratio vs BH mass for unobscured and obscured quasars including \w2246m05, following \cite{Wu_bhm}. The plotted data include \w2246m05\ (this work), J0100$+$2802 \citep{2015Natur.518..512W}, J1342$+$0928 \citep{2018Natur.553..473B}, QSOs at $z \gtrsim 6.4$ \citep{2017ApJ...849...91M}, QSOs at $z\sim 6$ \citep{2010ApJ...714..699W,2011ApJ...739...56D}, QSOs at $z \sim 4.8$ \citep{2011ApJ...730....7T}, SDSS QSOs at $1.5 < z < 5$ \citep{2011ApJS..194...45S}, \hotdogs\ at $z\sim 2$ \citep{Wu_bhm}, \hotdog\ at $z\sim 1$ \citep{2017ApJ...835..105R}, red QSOs \citep{2012MNRAS.427.2275B,2015MNRAS.447.3368B}, DOGs \citep{2011AJ....141..141M,2012AJ....143..125M}, and SMGs \citep{2008AJ....135.1968A}. Typical systematic uncertainties are shown by the error bars plotted at the lower left, with the errors for the broad \MgII\ line estimates in red and the errors for the broad Balmer line estimates in green.
}\label{fig:Eddington}
\end{figure*}

\subsection{$M_{\rm BH}$-$M_{\rm sph}$ Relation}

From a sample of five \hotdogs\ with $M_{\rm BH}$ measurements, \cite{Wu_bhm} found that the ratio of $M_{\rm BH}$ to the stellar mass in the spheroidal component of the host galaxy ($M_{\rm BH}$-$M_{\rm sph}$) in these systems is closer to the $M_{\rm BH}$-$M_{\rm sph}$ relation seen in local active galaxies \citep{2011ApJ...726...59B} than are the ratios seen in $z \sim 1.3$ quasars. 

We estimate the bulge mass in \w2246m05 from $K$ and $4.5 \mu$m photometry using the synthesized elliptical galaxy SED template from the GRASIL code \citep{1998ApJ...509..103S} to represent the spheroidal component, omitting the $3.6 \mu$m data point which may be significantly elevated by \Ha\ emission line (see Figure \ref{fig:SED}). In Figure \ref{fig:SED}, the brown solid line shows the SED of an elliptical galaxy at an age of 1.3 Gyr with $M_{\rm sph} = 1.1 \times 10^{12} M_{\sun}$. This can be considered as an upper limit of bulge mass of \w2246m05. This value is similar to the $\log (M_{\rm sph}/M_{\sun}) = 11.9$ value expected from the local $M_{\rm BH}$-$M_{\rm sph}$ relation \citep{2011ApJ...726...59B}:
\begin{eqnarray}
\log (M_{\rm BH}/M_{\sun}) = &&-3.34\pm1.91 \nonumber\\
&&+ (1.09\pm0.18) \times \log (M_{\rm sph}/M_{\sun}).
\end{eqnarray}
This suggests that \w2246m05\ has a similar $M_{\rm BH}$-$M_{\rm sph}$ relation to the \hotdogs\ shown in Figure 9 of \cite{Wu_bhm}.

\subsection{Eddington Ratio and Black Hole Accretion}

In \cite{Wu_bhm}, we argue that \hotdogs\ are on the high luminosity tail for a given $M_{\rm BH}$ with respect to SDSS QSOs because they achieve the highest accretion rates, implying that they are accreting material at the highest rates possible. To illustrate this, in Figure \ref{fig:Eddington}, we plot Eddington ratio ($\lambda_{\rm Edd} \equiv L_{\rm AGN} / L_{\rm Edd}$) vs. $M_{\rm BH}$. With $\log (M_{\rm BH}/M_{\sun}) = 9.6\pm0.4$, and $\log (L_{\rm bol}/L_{\sun}) = 14.6$, the Eddington ratio of \w2246m05\ is $\lambda_{\rm Edd} = 2.8_{-0.7}^{+0.9}$ (statistical uncertainty), the highest of all the \hotdogs\ for which we have so far obtained $M_{\rm BH}$ measurements \citep{Wu_bhm}, and putting the SMBH of \w2246m05\ well into the super-Eddington accretion region. The total uncertainty for the $\lambda_{\rm Edd}$, as shown by the dotted error bars in Figure \ref{fig:Eddington} for \w2246m05, is 0.4 dex, dominated by the systematic uncertainty of $M_{\rm BH}$. This systematic uncertainty also applies to the all objects at $z > 4$ in Figure \ref{fig:Eddington}.

Eddington ratios this large may attain a saturation level suggested by theoretical models \citep{1999ApJ...516..420W,2000PASJ...52..133W,2000PASJ...52..499M}. At high accretion rates, radiation pressure may dominate the accretion flow geometry, making the accretion disk ``slim'' \citep{1988ApJ...332..646A}. In these models, the fast radial transportation of mass in the accretion flow geometry can trap most photons, preventing them from escaping, and carrying them inward to the SMBH. This photon trapping accretion makes the radiation efficiency inversely proportional to the mass accretion rate. As a result, the luminosity of a super-Eddington accreting black hole may reach saturation at an apparent Eddington ratio of $\lambda_{\rm Edd} \sim 2$. The bolometric luminosity will increase much more slowly than the accretion rate, and the $M_{\rm BH}$ can increase much faster than the growth rate under the Eddington limit. In this scenario, the actual accretion rate of \w2246m05\ may be higher than the observed $\lambda_{\rm Edd}$ suggests.

Although they have similarly high luminosities, as shown in Figure \ref{fig:Eddington}, the \eEdd\ of \w2246m05\ may have reached saturation while the QSO J0100$+$2802 has not. The difference in the \eEdd\ of the sources is significant, considering only the statistical uncertainties from the measurements, but may not be when including systematic uncertainties in the $M_{\rm BH}$ estimate. We speculate that the difference between the sources, if real, is due to the much higher obscuration (thus more material to accrete) in \w2246m05\, or due to a different accreting geometry such as mass inflow from merging events. In this context, it is intriguing to note that recent ALMA observations of \w2246m05\ of dust continuum emission at rest $212 \mu$m reveal bridges of dusty, metal-enriched material connecting three companions to the central galaxy \citep{diaz-santos2017}, implying that a multiple merger is in progress. \cite{diaz-santos2017} estimate that accretion rates of as high as $\dot{M} \sim 900\, M_{\sun}\ {\rm yr}^{-1}$ onto the central core of \w2246m05\ could be underway - sufficient to power the high luminosity of \w2246m05\, which has a BH accretion rate of $24\, M_{\sun}\ {\rm yr}^{-1}$ if a radiation efficiency of 0.1 is assumed.  

\section{Conclusion}\label{sec:conclusions}
We report observations of broad \MgII\ and \CIV\ lines in the $z=4.601$ source \w2246m05, the most luminous galaxy known, providing clear evidence for the presence of an AGN in the system. The FWHM of \MgII\ is 3300\,\kms. Using the well-determined bolometric luminosity and an AGN template to estimate the 3000\,\AA\ continuum luminosity, we measure the black hole mass $\log (M_{\rm BH}/M_{\sun}) = 9.6 \pm 0.4$. The broad (5900 \kms) \CIV\ line is significantly blueshifted, (by 3300\,\kms), and we estimate the corresponding $M_{\rm BH}$ for this line using an empirically calibrated correction for the FWHM based on the blueshift. This method yields $\log (M_{\rm BH}/M_{\sun}) = 9.6 \pm 0.4$, in good agreement with the \MgII\ estimate.

We reevaluate the bolometric luminosity of \w2246m05\, considering the possible contribution of a nearby foreground galaxy. SCUBA2 450\,\micron\ observations show that \w2246m05\ dominates the far-IR flux. Using an ALMA 252 GHz continuum map, we estimate the contribution from the foreground galaxy is $<$10\%. The updated estimate of $L_{\rm bol}$, based on power-law interpolation of the well sampled SED, is $3.6\times 10^{14}\,L_{\sun}$. The SED shows a dip near rest $10 \mu$m suggestive of silicate absorption. 

The Eddington ratio in \w2246m05\ is 2.8. Theoretical arguments suggest the luminosity may be saturating in this super-Eddington regime, and be insensitive to the mass accretion rate. In this scenario, the $M_{\rm BH}$ growth rate may hence exceed the apparent accretion rate derived from the observed luminosity.

\acknowledgments
The authors gratefully acknowledge the anonymous referee for valuable comments and suggestions that improved the paper.
C.-W.T. would like to thank Jian-Min Wang for the valuable discussions. 
This material is based upon work supported by the National Aeronautics and Space Administration under Proposal No. 13-ADAP13-0092 issued through the Astrophysics Data Analysis Program. 
H.D.J. is supported by Basic Science Research Program through the National Research Foundation of Korea (NRF) funded by the Ministry of Education (NRF-2017R1A6A3A04005158).
J.W. is supported by the National Key Program for Science and Technology Research and Development of China (grant 2016YFA0400702) and Project 11673029 supported by NSFC.
RJA was supported by FONDECYT grant number 1151408.
T.D.-S. acknowledges support from ALMA-CONICYT project 31130005 and FONDECYT regular project 1151239.
This publication makes use of data obtained at the W.M. Keck Observatory, which is operated as a scientific partnership among Caltech, the University of California and NASA. The Keck Observatory was made possible by the generous financial support of the W.M. Keck Foundation. 
The authors wish to recognize and acknowledge the very significant cultural role and reverence that the summit of Mauna Kea has always had within the indigenous Hawaiian community. We are most fortunate to have the opportunity to conduct observations from this mountain.
This publication makes use of data products from the {\it Wide-field Infrared Survey Explorer}, which is a joint project of the University of California, Los Angeles, and the Jet Propulsion Laboratory, California Institute of Technology, funded by the National Aeronautics and Space Administration. 
This research has made use of the NASA/ IPAC Infrared Science Archive, which is operated by the Jet Propulsion Laboratory, California Institute of Technology, under contract with the National Aeronautics and Space Administration.

{\textit{Facilities:}} \facility{\textit{Keck:I (OSIRIS, LRIS)}}, \facility{\textit{Herschel Space Telescope}}, \facility{\textit{Wide-field Infrared Survey Explorer}}, \facility{\textit{JCMT}} \\


\end{document}